\documentclass[a4paper,fleqn]{cas-dc}

\renewcommand{\printorcid}{}  % supress ORCIDs

\usepackage[numbers]{natbib}

\def\tsc#1{\csdef{#1}{\textsc{\lowercase{#1}}\xspace}}
\tsc{WGM}
\tsc{QE}
\tsc{EP}
\tsc{PMS}
\tsc{BEC}
\tsc{DE}
\begin{document}
\let\WriteBookmarks\relax
\def\floatpagepagefraction{1}
\def\textpagefraction{.001}
\shorttitle{Relative variations of nonlinear elastic moduli in polystyrene-based nanocomposites}
\shortauthors{A.V. Belashov et~al.}

\title [mode = title]{Relative variations of nonlinear elastic moduli in polystyrene-based nanocomposites}                      
%\tnotemark[1,2]

%\tnotetext[1]{This document is the results of the research
%   project funded by the National Science Foundation.}

%\tnotetext[2]{The second title footnote which is a longer text matter
%%   another line in the footnotes area of the first page.}

\author[1]{A.V. Belashov}[]
%\cormark[1]
%\fnmark[1]
%\ead{cvr_1@tug.org.in}
%\ead[url]{www.cvr.cc, cvr@sayahna.org}

%\credit{Conceptualization of this study, Methodology, Software}

\address[1]{Ioffe Institute, 26, Polytekhnicheskaya, St.Petersburg, 194021, Russia}

\author[1]{Y.M. Beltukov}[]

\author[2]{O.A. Moskalyuk}[]
%\fnmark[2]
%\ead{cvr3@sayahna.org}
%\ead[URL]{www.sayahna.org}

%\credit{Data curation, Writing - Original draft preparation}

\address[2]{St.Petersburg State University of Industrial Technologies and Design, 18, Bolshaya Morskaya, St. Petersburg, 191186, Russia}

\author[1]
{I.V. Semenova}
\cormark[1]
%\fnmark[1,3]
%\ead{irina.semenova@mail.ioffe.ru}
%\ead[URL]{www.stmdocs.in}

%\address[3]{STM Document Engineering Pvt Ltd., Mepukada,
%    Malayinkil, Trivandrum 695571, India}

\cortext[cor1]{Corresponding author:
irina.semenova@mail.ioffe.ru (I.V. Semenova)}
%\cortext[cor2]{Principal corresponding author}
%\fntext[fn1]{This is the first author footnote. but is common to third
%  author as well.}
%\fntext[fn2]{Another author footnote, this is a very long footnote and
%  it should be a really long footnote. But this footnote is not yet
%  sufficiently long enough to make two lines of footnote text.}

%\nonumnote{This note has no numbers. In this work we demonstrate $a_b$
%  the formation Y\_1 of a new type of polariton on the interface
%  between a cuprous oxide slab and a polystyrene micro-sphere placed
%  on the slab.
%  }

\begin{abstract}
In this paper we apply the methodology based on the analysis of changes in acoustic wave velocities under static stress for measurements of the third-order elastic moduli in three polystyrene-based nanocomposites with different fillers: SiO${}_2$ particles, halloysite natural tubules, and carbon black particles. The samples were fabricated by the same technology and our data provide information on relative changes of nonlinear properties of the composites caused by addition of the fillers. The data obtained for composites are compared with that for commercial grade polystyrene.
The substantial variations of the nonlinear elastic moduli for composites with different types of fillers are demonstrated and analyzed. The results are in a qualitative agreement with theoretical predictions. 
\end{abstract}

% \begin{graphicalabstract}
% \includegraphics{figs/grabs.pdf}
% \end{graphicalabstract}

% \begin{highlights}
% \item Research highlights item 1
% \item Research highlights item 2
% \item Research highlights item 3
% \end{highlights}


\maketitle

\section{Introduction}

Due to the current progress in engineering functional nano-, micro- and macro-composites are widely used in various industries, in aerospace, automotive technologies, and pipe line transportation, in particular \cite{aerospace-comp2018,automotive,gibson2010,multifunc2016}. Rapidly growing applications of composite materials stimulate analysis of various aspects of their mechanical behavior. However, the composite response to intensive dynamic loading is often hard to predict since it may depend drastically upon matrix and filler characteristics, filler distribution in the matrix and on resulting mechanical characteristics of the composite. The reliable prediction of novel composite materials behavior under dynamic loading is one of the most demanded problems of the modern condensed matter physics from the viewpoint of both theoretical description and applications.

Nonlinear elastic properties of materials are increasingly found to be of high importance in the description of material response to various loads. A number of researchers have suggested approaches to define nonlinear elasticity using third-order elastic moduli, or nonlinearity parameters, that are applied as measures to evaluate the deviation of a stress-strain relationship from linear behavior. The ones used most frequently for the description of nonlinear elasticity in solids are those introduced by Landau and Lifshitz \cite{Landau}, Murnaghan \cite{Murnaghan}  and Thurston and Brugger \cite{Brugger1964}. The sets of nonlinear elastic moduli defined in these models are in fact closely related and can be calculated one from another.

The third-order elastic moduli and their linear combinations were already demonstrated to be informative for the prediction of fatigue damage, for description of thermoelastic properties of crystalline solids, acoustic radiation stress, radiation-induced static strains, creep, thermal aging, wave processes, etc.
In general nonlinear parameters were demonstrated to be more sensitive to structural changes in the material than linear ones.

In Murnaghan's approximation the nonlinear elastic behavior of an isotropic solid material is described by three nonlinear, third-order moduli ($l,m,n$) and two linear, 2nd order, Lam\'e constants ($\lambda, \mu$). First measurements of these constants were performed by Hughes and Kelly, based on the simplified  Murnaghan's theory, see \cite{Hughes-Kelly1953} for details. In brief, the methodology is based on the acousto-elastic effect and applies the analysis of the dependence of  velocity of longitudinal and shear ultrasound waves in the sample upon the applied static stress, providing data for determination of nonlinear elastic moduli of the material. This technique was then widely applied the measurements of nonlinear elastic moduli of various materials, see e.g. \cite{egle1976,winkler1996,renaud2016} and is still one of the most frequently used approaches. 

Other approaches developed to assess nonlinear elastic properties of materials utilize dynamic acousto-elasticity \cite{DAE_Renaud,DAE_Payan}, second harmonic generation \cite{SHG}, Brillouin scattering \cite{Moduli_polycarb}, coda wave interferometry \cite{Coda_wave}, strain solitary \cite{Soliton_moduli}, Lamb  \cite{Lamb-waves} and Rayleigh \cite{Rayleigh-waves} waves.
However, no standard measurement procedure has been approved by now.

There are many theoretical and numerical studies on the linear elastic moduli of nanocomposites \cite{eshelby57, HalpinKardos, Adnan-2007, Thostenson-2003}. However, theoretical models of the nonlinear elastic properties of composite materials are still in an active development \cite{sevostianov2001,tsvelodub2000,tsvelodub2004,giordano2009, ColomboGiordano, giordano2017}. A theory, which takes into account  nonlinear elastic properties of both  the matrix and the filler was developed recently for the case of spherical inclusions \cite{Semenov_Beltukov2020}. The experimental validation of theoretical predictions is highly desirable, however by now measurements of nonlinear elastic properties of composites are very rare. And to the best of our knowledge no data on that for polymeric composites has been published as yet.

%It is profound to obtain experimental results, which can be compared with the theoretical predictions. 

In this paper we apply the methodology based on the analysis of changes in acoustic wave velocities under static stress for measurements of the third-order elastic moduli in three polystyrene-based nanocomposites with different fillers. The samples were fabricated by the same technology and our data provide information on relative changes of nonlinear properties of the composites provided by addition of the fillers. The results obtained for composites are compared with those for commercial grade pure polystyrene. The correlation of experimental data with theoretical predictions is analyzed.

\section{Methodology for evaluation of the third-order elastic moduli}

%\subsection{Acoustoelastic effect and elastic moduli}

According to Hughes and Kelly \cite{Hughes-Kelly1953} in the case when no shear deformation in the sample takes place and strain tensor is diagonal, velocities of longitudinal and shear ultrasonic  waves propagating in the sample along the $x$ direction can be expressed in the form:
\begin{align}
\label{set1.1}
\!\!\! \rho_0 V_x^2 &= \lambda + 2\mu + (2l + \lambda)\operatorname{Tr}{\varepsilon} + (4m + 4\lambda + 10\mu)\varepsilon_{xx}, \\
\label{set1.2}
\!\!\! \rho_0 V_y^2 &= \mu + (\lambda + m)\operatorname{Tr}{\varepsilon} + 4\mu\varepsilon_{xx} + 2\mu\varepsilon_{yy} - \frac{n}{2}\varepsilon_{zz}, \\
\label{set1.3}
\!\!\! \rho_0 V_z^2 &= \mu + (\lambda + m)\operatorname{Tr}{\varepsilon} + 4\mu\varepsilon_{xx} + 2\mu\varepsilon_{zz} - \frac{n}{2}\varepsilon_{yy},
\end{align}
where $\rho_0$ is the undeformed sample density, $V_x$ is longitudinal wave velocity, and $V_y$, $V_z$ are velocities of shear waves  polarized parallel
and perpendicular to the uniaxial stress and traveling
perpendicular to the stress axis.  
Assume the static uniaxial stress $T$ is applied along the $y$ axis, then three non-zero components of the strain tensor can be written as: $\varepsilon_{yy} = -T/E$, $\varepsilon_{xx} = \varepsilon_{zz} = \nu T/E$, where Young's modulus $E=\mu(3\lambda + 2\mu)/(\lambda + \mu)$  and Poisson coefficient $\nu = \lambda/(2\lambda + 2\mu)$. 
Introduce the effective longitudinal and shear moduli: $M_x = V_x^2 \rho_0$, $G_y = V_y^2 \rho_0$, $G_z = V_z^2 \rho_0$ and write the set of equations (\ref{set1.1})--(\ref{set1.3}) in the form:
\begin{align}
\label{M-modulus}
\!\!\! M_x &= \lambda + 2\mu + \alpha_x T,&
\alpha_x &= -\frac{2l - \frac{\lambda}{\mu}(2m + \lambda + 2\mu)}{3\lambda + 2\mu},\\
\label{Gy-modulus}
\!\!\! G_y &= \mu + \alpha_y T,&
\alpha_y &= -\frac{m + \frac{\lambda }{4\mu}n + 2\lambda + 2\mu}{3\lambda + 2\mu},\\
\label{Gz-modulus}
\!\!\! G_z &= \mu + \alpha_z T,&
\alpha_z &= -\frac{m - \frac{\lambda+\mu}{2\mu}n - \lambda}{3\lambda + 2\mu},
\end{align}
where $\alpha_x$, $\alpha_y$ and $\alpha_z$ are dimensionless slope coefficients of the dependencies of corresponding moduli as functions of applied transverse stress. The linear elastic moduli $\lambda$, $\mu$ can be found from the longitudinal and shear velocities at zero strain, while nonlinear ones $l$, $m$, $n$ can be calculated from the slope coefficients $\alpha_x$, $\alpha_y$ and $\alpha_z$. Thus, the whole set of measurements provides data on both the linear and nonlinear elastic moduli of the specimen material.
In the general case we have:
\begin{gather}
\!\!\! l = -\frac{3 \lambda +2 \mu }{2} \alpha _x-\frac{\lambda  (\lambda +\mu ) }{\mu}(1+2 \alpha   _y)+\frac{\lambda ^2 }{2 \mu }(1-2 \alpha_z),\\
\!\!\! m = -2 (\lambda +\mu ) \left(1+\alpha_y\right)+\lambda  \left(1-\alpha_z\right),\\
\!\!\! n = -4 \mu  \left(1+\alpha _y-\alpha _z\right). 
\end{gather}

As a simplified alternative to the analysis of the set of third-order moduli 
nonlinear elastic properties of materials can be evaluated using the nonlinearity	 parameter $\beta$ comprising a combination of
Murnaghan's and Lame's elastic moduli (see e.g. \cite{Ostr_Johnson2001,Coda_wave}): 	
\begin{equation}\label{beta}
\beta = 3/2 + (l+2m)/(\lambda + 2\mu).
\end{equation}
Note that the $n$ modulus is omitted in the parameter $\beta$. 
Conversely, it can be easily shown from equation (\ref{M-modulus}) that if the Poisson coefficient $\nu$ equals to 1/3 (it holds for most of glassy polymers), then a slightly different parameter $\gamma$ can be introduced and obtained just from measurements of longitudinal wave velocity as a function of the applied uniaxial stress:   
\begin{equation}\label{l-2m}
\gamma = (l-2m)/(\lambda + 2\mu) = 1-\alpha_x.
\end{equation}
%\begin{equation}\label{l-2m}
%l-2m = %(1-\alpha_x)(\lambd%a + 2\mu)
%\end{equation}

As it was recently shown  \cite{Semenov_Beltukov2020} in nanocomposites on the base of glassy polymers, such as PS, PMMA or polycarbonate $l$ modulus demonstrates the most profound changes, while $m$ and $n$ moduli show the least variations. That is why in many practical cases the estimation of $\gamma$ value obtained from measurements of just longitudinal wave velocity as a function of applied stress can provide data for rough tentative estimation of potential changes in nonlinear properties of a nanocomposite.

\section{Materials and experimental setup}

\subsection{Materials}

The grained 585 polystyrene (Nizhnekamskneftekhim, Russia)  was used as a polymer matrix for composite samples. The following materials were applied as nanofillers: silicon dioxide (SiO${}_2$) particles Aerosil R812 modified by silazane (Evonic Industries, Germany); Carbon Black (CB) particles P-805E (Ivanovskiy tekuglerod and rubber, Russia) and Halloysite Natural Tubules (HNT) (NaturalNano Inc., USA). The typical sizes of filler particles were as follows: SiO${}_2$ $\sim$ 7 nm in diameter, CB $\sim$ 80 nm in diameter, HNT $\sim$ 100 nm in diameter, 0.5 -- 1.2 {\textmu}m long.
Spherical particles, SiO$_2$ and CB, were introduced to the PS-matrix at the concentrations of up to 20\% weight and halloysite tubules of up to 10\% weight. The chosen filler concentrations were shown to provide the most essential changes of linear elastic properties  without significant  agglomeration \cite{nanocomp2020}.  

Composite samples were fabricated by melt technology in the form of plates $1\times10\times50$ mm$^3$, see \cite{nanocomp2017,nanocomp2019} for fabrication details. Similar plates were fabricated by the same technology from pure polystyrene and used for reference.  

\begin{figure}
\includegraphics[width=\linewidth]{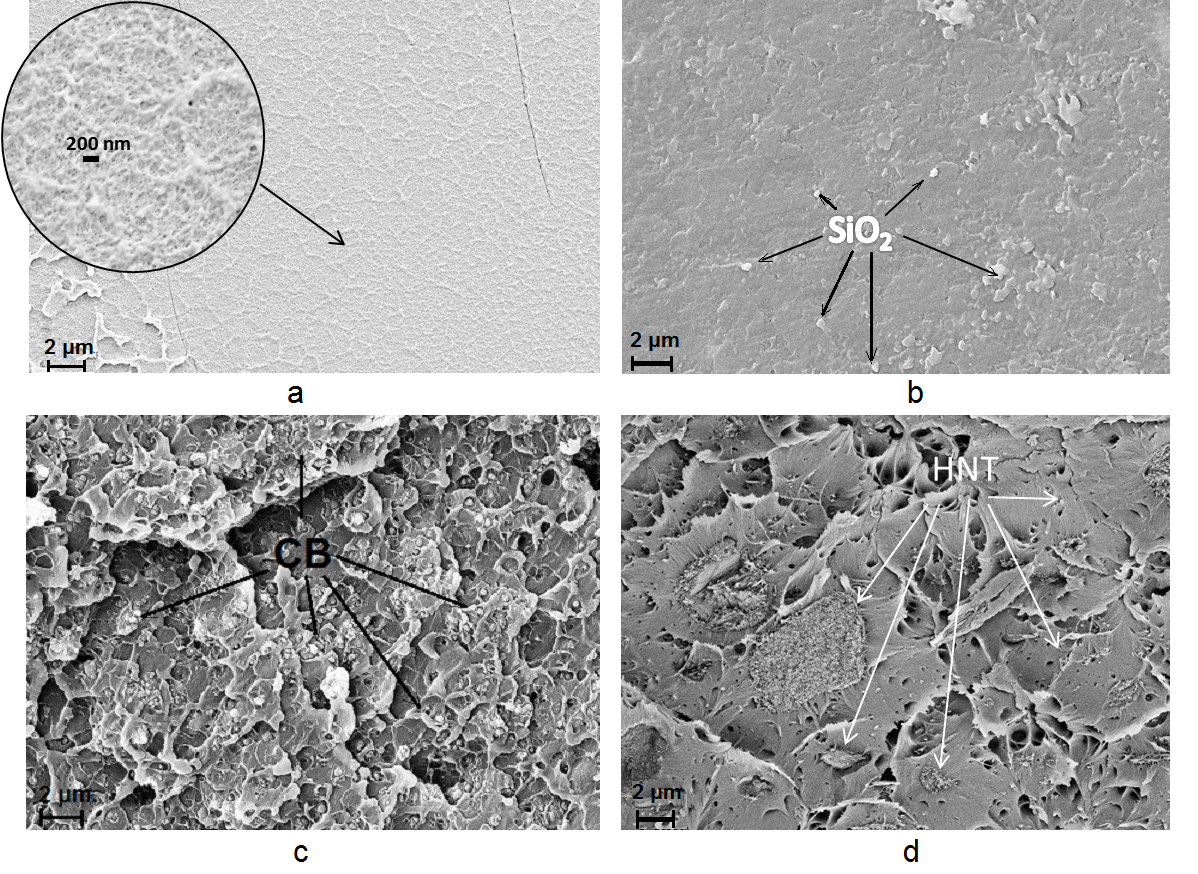} % Here is how to import EPS art
\caption{\label{microphoto} Microphotographs of cryo-cleaved surfaces of samples.}
\end{figure}

As known one of the critical requirements for good quality polymer nanocomposite is a well dispersion of filler particles in the polymer matrix and lack of agglomeration \cite{bhattacharya2016}. The particle distribution in the PS matrix was controlled by observation of micrographs of cryo-cleaved surfaces of the composite samples using a Carl Zeiss Supra-55 scanning electron microscope. Figure \ref{microphoto} presents representative microphotographs of the composite samples.

The influence of fillers on the static mechanical properties of nanocomposites in tension was studied using Instron 5940 universal testing system (USA). Basing on the tensile test data, the strength at break $\sigma{_b}$, strain at break $\epsilon{_b}$, and static tensile elastic modulus $E$ have been determined, see Table~\ref{comp-parameters}.
As can be seen from these data the introduction of more rigid particles leads to a more profound increase in the elastic modulus. The elastic modulus rose from 1.6 GPa for pure PS to 3.0 GPa for composites with 20\% CB. Also the CB particles provided high strength and strain at break of composites, with strength at the break being even higher than that of pure PS. 
Thus the chosen concentrations of nanofillers on one hand provided a noticeable change of material elasticity but on the other hand did not cause its substantial agglomeration.

\begin{table}
\centering
\caption{\label{comp-parameters}Data from static tensile tests of PS-based composites.}
%\begin{ruledtabular}
\begin{tabular}{cccc}
\textbf{} & \textbf{Strength} & \textbf{Tensile} & \textbf{Strain}\\
\textbf{Sample} & \textbf{at break,} & \textbf{modulus,} & \textbf{at break,}\\
    \textbf{} & \textbf{$\sigma{_b}$, MPa} & \textbf {$E$, GPa} & \textbf{$\epsilon{_b}$, \%}\\
\hline
PS pure           & 5 6$\pm$ 1 & 1.6 $\pm$ 0.1 & 5.6 $\pm$ 0.2\\
PS+20\% SiO${}_2$ & 31 $\pm$ 3 & 1.8 $\pm$ 0.2 & 2.2 $\pm$ 0.2\\
PS+10\% HNT       & 49 $\pm$ 6 & 2.0 $\pm$ 0.1 & 2.8 $\pm$ 0.5\\
%PS+10\% CB       & 74 $\pm$ 2 & 2.8 $\pm$ 0.1 & 3.1 $\pm$ 0.2\\
PS+20\% CB        & 65 $\pm$ 5 & 3.0 $\pm$ 0.2 & 2.1 $\pm$ 0.6\\
\end{tabular}
%\end{ruledtabular}
\end{table}

\subsection{Experimental setup}

\begin{figure}
\includegraphics[width=7 cm]{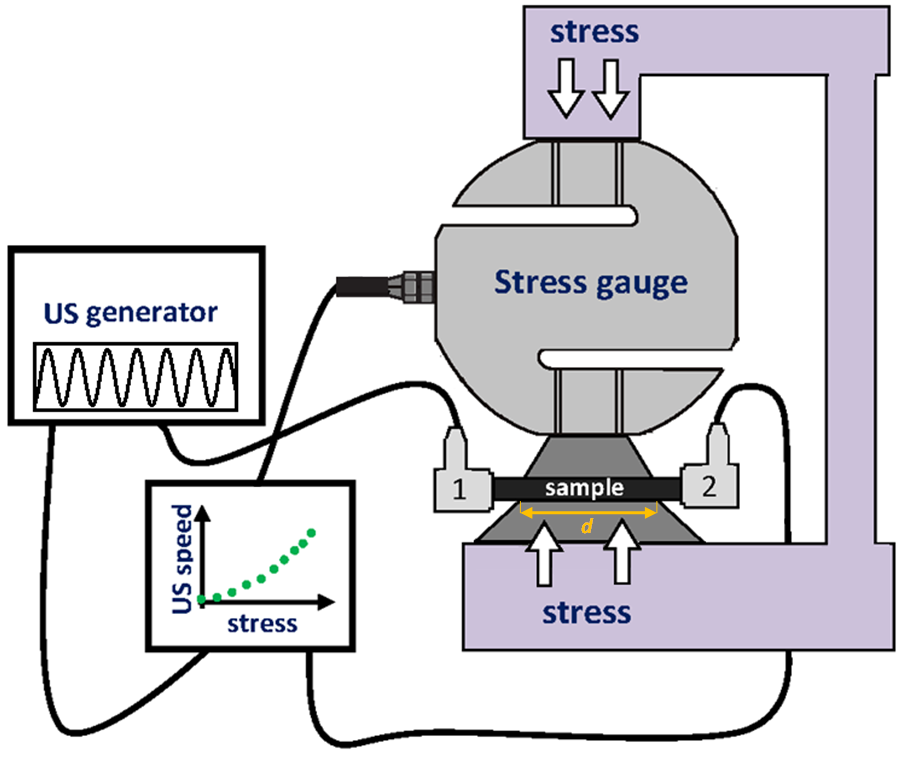}% Here is how to import EPS art
\caption{\label{Scheme_US} Experimental setup for measurements of ultrasound velocity as a function of applied stress.}
\end{figure}

The analysis of nonlinear elastic properties of fabricated nanocomposite samples was performed through measurements of changes in  velocities of longitudinal and shear ultrasonic waves depending upon the applied static stress. The schematic of the experimental setup is shown in Fig.~\ref{Scheme_US}.
Each specimen used in measurements was formed from the three fabricated plates of a nanocomposite,  adhesively bonded by the ethylcyanoacrylate adhesive Superglue. The specimens had a bar shape with the length of about 5 cm. A specimen and a stress gauge were clamped between the jaws of the stress unit. Special metallic vice grips indicated by gray color in Fig.~\ref{Scheme_US} provided uniform stress on a certain area of the specimen of the length $d$. Two piezoelectric transducers were applied to the end faces of the specimen and were used for generation and detection of longitudinal (piezoelectric transducers P121 by Amati Acoustics, Russia) and shear  (piezoelectric transducers V154-RB by Olympus, USA) waves. The input signals were provided by the AM300 Dual Arbitrary Generator (Rohde \& Schwarz), output signals were recorded by RTB2002 digital oscilloscope (Rohde \& Schwarz). 

\subsection{Measurement procedure}

In the first set of measurements longitudinal and shear wave velocities were obtained for each of the specimens at zero applied stress. The pulsed longitudinal/shear waves were generated at the input end face of the specimen and detected at the output one. Ultrasonic wave velocity was estimated as a ratio of the specimen length and time delay between the fronts of the input and output pulses. The obtained values of longitudinal and shear wave velocities in the specimens are summarized in Table \ref{velocities}.

\begin{table}
\centering
\caption{\label{velocities}Measured longitudinal and shear ultrasonic wave velocities.}
%\begin{ruledtabular}
\begin{tabular}{ccc}
\textbf{Specimen} & \textbf{Longitudinal,} & \textbf{Shear,}\\
    \textbf{} & \textbf{m/s} & \textbf {m/s} \\
\hline
PS commercial & 2316 $ \pm$ 4 & 1166 $ \pm$ 6 \\
PS pure & 2312 $ \pm$ 4 & 1170 $ \pm$ 6 \\
PS+20\% SiO${}_2$ & 2348 $\pm$ 4 & 1158 $ \pm$ 6\\
PS+10\% HNT & 2357 $ \pm$ 4 & 1199 $ \pm$ 6\\
%PS+10\% CB & 74$\pm$ 2 & 2.8$\pm$ 0.1 & 3.1 $\pm$ 0.2\\
PS+20\% CB & 2456 $ \pm$ 4 & 1215 $ \pm$ 6\\
\end{tabular}
%\end{ruledtabular}
\end{table}

A different approach was utilized for measurements of velocity variations as a function of the applied static stress.
In this set of experiments, sine ultrasonic waves with the frequency of 2.25 MHz were applied to the input end face of the specimen and detected at the output end face. Shifts of the sine wave at stepwise increase of the applied static stress were recorded providing data on changes of wave velocity.   Relatively small changes of the applied stress (up to 15 MPa) did not cause significant variations of longitudinal/shear wave velocity and sine wave shifts did not exceed its period. The advantage of  sinusoidal shape of the ultrasonic wave in this case allowed us to choose a single frequency and to observe a signal of a constant shape which did not vary with the applied stress. 

\begin{figure}
\includegraphics[width=8.2 cm]{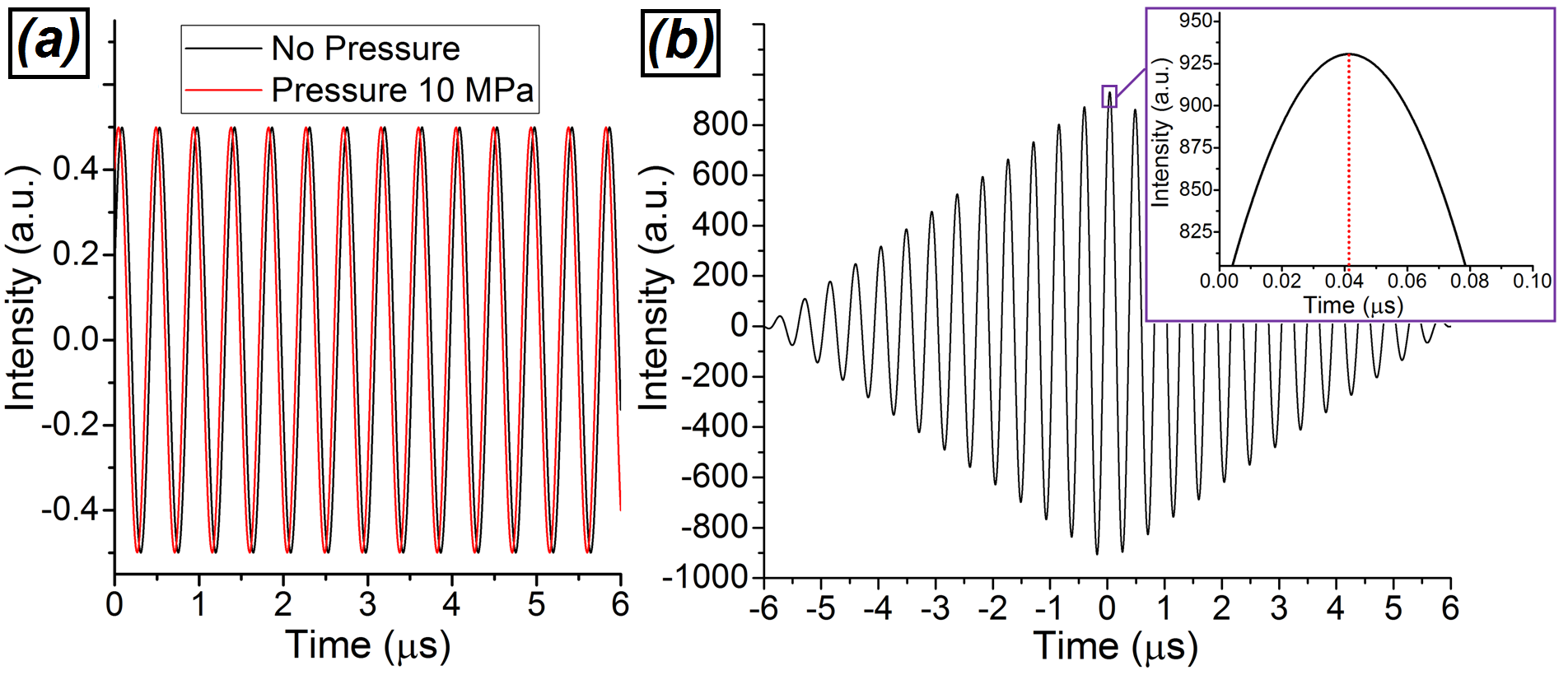}% Here is how to import EPS art
\caption{\label{Experiment_sample} (a) Typical recorded sinusoidal signals obtained for the specimen of commercial grade PS at zero stress and at T = 10 MPa. (b) Convolution between the two signals shown in (a) allows for estimation of temporal shift between the two signals as $\Delta t_P = 0.0412$ {\textmu}s.}
\end{figure}

Calculation of the sine wave shift induced by the applied stress can be performed by convolution of the signal obtained at zero stress $S_0(t)$ and that  recorded at higher stress $S_P(t)$: $R_P(t) = S_P(t) \ast S_0(t)$. If two sinusoidal signals are shifted from each other at time $\Delta t_P$ (Fig.~\ref{Experiment_sample}(a)), global maxima of their convolution $C_P(t)$ will be shifted from $t=0$ at a distance $\Delta t_P$ as shown in Fig.~\ref{Experiment_sample}(b). Since the time delay $\Delta t_P = \frac{d}{V_P} - \frac{d}{V_0}$ is a result of wave velocity change from the initial value $V_0$ to the final one $V_P$ and occurs on a specimen length $d$ where stress had been applied, the ultrasonic wave velocity at stress $P$ can be calculated as: 
\begin{equation}\label{VelCalc}
V_P = \frac{d}{\Delta t_P V_P + d} V_0.
\end{equation}

\section{Results and discussion}

\begin{figure}
\includegraphics[width=8.2 cm]{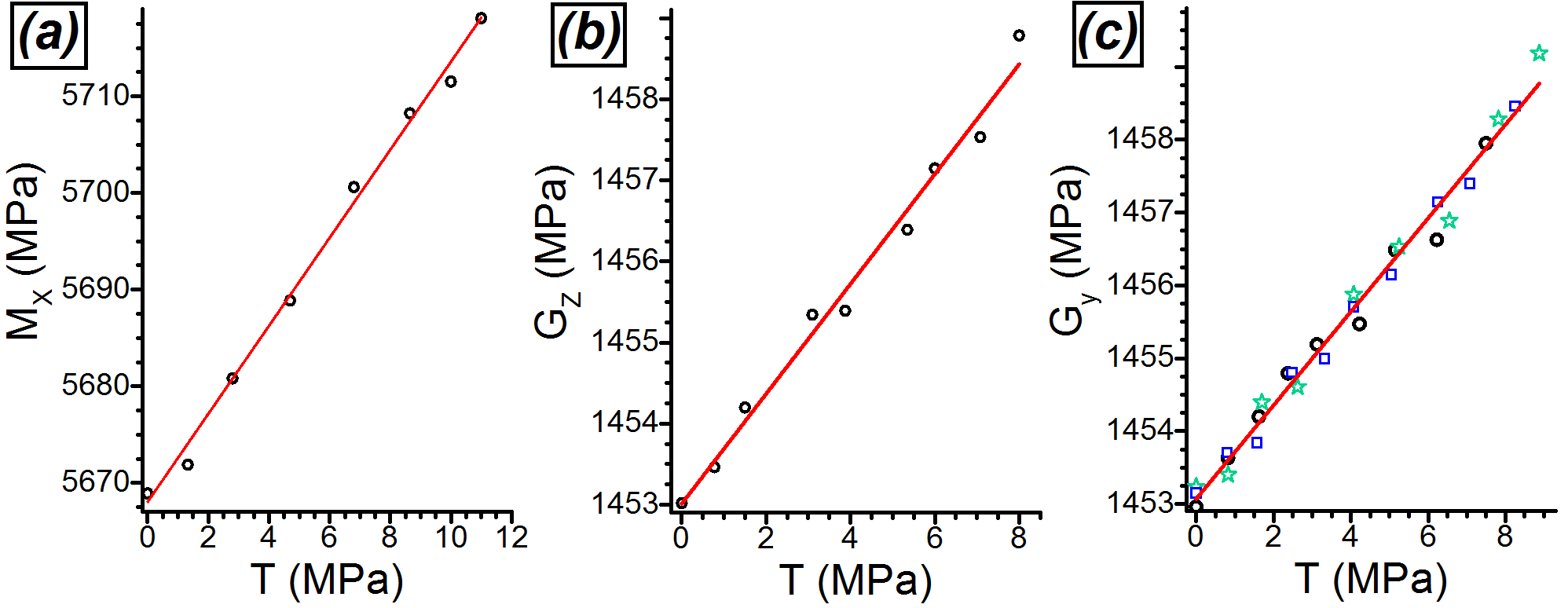}% Here is how to import EPS art
\caption{\label{OlgaSampleDetails} Example of measured dependencies of effective longitudinal (a) and shear (b,c) moduli on applied stress T for the PS pure specimen. Figure (c) also demonstrates the reproducibility of measurements for a single investigated specimen, with colored symbols indicating different sets of measurements.}
\end{figure}

The described methodology was used for measurements of wave velocity as a function of applied static stress $T$ for three types of waves: 1) longitudinal waves $V_x(T)$, 2) shear waves orthogonal to the direction of applied stress $V_z(T)$ and 3) shear waves parallel to the direction of applied stress $V_y(T)$. Taking into account the PS density ($\rho = 1060$ kg/m${}^3$) or the corresponding composite density, effective longitudinal and shear moduli at several stress values were calculated for each specimen. An example of the complete set of these dependencies obtained for the sandwich specimen made of fabricated plates of pure polystyrene (PS pure) is shown in Fig.~\ref{OlgaSampleDetails}. The obtained sets of data were used for calculations of the second- and third-order moduli of specimens made of pure polystyrene and three PS-based nanocomposites: PS + 20\% SiO${}_2$, PS + 10\% HNT and PS + 20\% CB. For comparison measurements were also performed in a bulk non-layered specimen made of a commercial grade pure polystyrene.  The results obtained are summarized in Table \ref{tab:table3}. The data obtained by Hughes and Kelly \cite{Hughes-Kelly1953} for pure polystyrene are shown for comparison.

The following conclusions can be made from the analysis of data in
Table \ref{tab:table3}. First of all the elastic characteristics of the modern commercial grade polystyrene differ noticeably from those of the one examined in 1950-s by Hughes and Kelly \cite{Hughes-Kelly1953}. That concerns both the linear and nonlinear parameters. Parameters of the sandwich made of adhesively bonded plates of pure polystyrene fabricated by melt technology slightly differ from those of commercial grade bulk specimen, that can be due to several factors, among which are the difference in polymer structure and fabrication process and the influence of adhesive layers. 
Sheets of commercial grade PS were fabricated by the extrusion technology followed by cold rolling. While laboratory samples were fabricated by melt injection into the heated mold under pressure followed by self-cooling. These technologies cause differences in the formed polymer structure that result in variations of the nonlinear elastic moduli, which are more sensitive to structural changes than the linear ones.
Notably, the linear elastic moduli of the commercial and laboratory polystyrene samples demonstrate almost no difference. 
We have to mention also that differences in both linear and nonlinear elastic parameters of these two materials are essentially lower than those of the modern commercial grade PS and the one used in \cite{Hughes-Kelly1953}. This observation also supports our earlier experimental evidence that layers of ethylcyanoacrylate Superglue adhesive do not affect markedly the resulting material nonlinearity, that was demonstrated in terms of parameters of bulk nonlinear strain solitary waves propagating in layered polymeric waveguides \cite{jap2008,apl2018}.

With this result in mind let us compare elastic parameters of nanocomposites with those of pure polystyrene sandwich fabricated by the same technology (PS pure in Table \ref{tab:table3}). As can be readily seen from the Table all the three nanofillers cause noticeable rise of the second order Lame moduli $\lambda$ and $\mu$ except for the latter for PS with SiO${}_2$ particles. The Young's modulus $E$ calculated from these data comprises 3.85 GPa for PS pure and 
% standard density:
%3.80, 4.05 and 4.18 GPa 
% calculated density:
4.23, 4.29 and 4.58 GPa 
for nanocomposites with SiO${}_2$, HNT and CB particles respectively. Thus in terms of linear elastic properties the addition of 10\% HNT and 20\% CB to the polystyrene matrix provided efficient reinforcement of the material. The Young's modulus $E$ of composites obtained by ultrasonic measurements exceeded the static one (Table \ref{comp-parameters}) by approximately 2 GPa. It corresponds to a typical frequency dependence of elastic moduli of polymer materials \cite{Yadav-2020}.

Changes in the nonlinear, third-order moduli $l$, $m$ and $n$, demonstrate more complex behavior. In general for all the composite samples  changes in the nonlinear moduli were more profound than changes in the linear, 2-nd order moduli $\lambda$ and $\mu$. 
% standard density:
%Among the nonlinear moduli the $l$ modulus demonstrated prominent variations in all the composites with the maximal change of about 65\% observed in the CB-containing one. The $n$ modulus showed high relative variation, of about 80\%, only in PS+HNT samples, while variations of the $m$ modulus were not so high and reached 40\% in PS+HNT nanocomposite. 
% calculated density:
Among the nonlinear moduli the $l$ modulus demonstrated  prominent variations in all the composites with the maximal change of about 84 \% observed in the CB-containing one. The $n$ modulus showed high relative variation, of about 91 \% only in PS+HNT samples, while variations of the $m$ modulus were not so high and reached 46\% 
in PS+HNT and PS+CB nanocomposites. 

The values of nonlinearity parameter $\beta$ calculated using equation (\ref{beta}) show comparatively small difference between the laboratory and commercial grade polystyrene samples and, surprisingly, a very small difference between the pure laboratory PS and PS+HNT composite as compared with higher difference of those and PS+SiO${}_2$ composite. The parameter $\gamma$ (Eq. (\ref{l-2m})) demonstrates  more straightforward behavior in these composites. The change of $\gamma$ values in HNT-containing composite is higher than that in SiO${}_2$-containing one. Meanwhile, both $\beta$ and $\gamma$ absolute values show profound rise in the CB-containing composite as compared with pure PS.

%The analysis of  $l-2m$ values demonstrates substantial changes. In particular, while for PS pure $l-2m$ equals to -18.96, for the composites it comprises -29.15 for PS+SiO${}_2$, -3.53 for PS+HNT and -40.29 for PS+CB.    

The obtained linear and nonlinear moduli of PS + 20\% SiO${}_2$ and PS + 20\% CB can be compared with the developed theory of nonlinear elastic properties of composites with spherical inclusions \cite{Semenov_Beltukov2020}. This theory predicts the effective values of elastic moduli $\lambda, \mu, l, m, n$ using the known values of these moduli for the nanoparticles and the surrounding matrix. The values for the matrix were taken from our experiment for pure PS (see Table~\ref{tab:table3}). To estimate the elastic properties of SiO${}_2$ nanoparticles, we used the properties of bulk amorphous silica: Young's modulus $E=72$ GPa and shear modulus $\mu=31$ GPa \cite{Pabst-2013}. For CB nanoparticles we took the Young's modulus $E=80$ GPa and Poisson ratio $\nu = 0.3$ \cite{Jean-2011}. In both cases, the nanoparticles are much stiffer than the polymer matrix. Thus, the influence of nonlinear moduli of nanoparticles to effective nonlinear moduli of nanocomposite is negligible. Indeed, for any macroscopic deformation of the composite, the deformation of nanoparticles is much smaller than that of the surrounding matrix. To estimate the volume fraction of nanoparticles we used the density $\rho = 2200$ kg/m${}^3$ and $\rho=1900$ kg/m${}^3$ for SiO${}_2$ and CB respectively. The resulting values are presented in Table~\ref{tab:table3}. Most of the theoretical predictions give values relatively close to the experimental ones. However, the experiment demonstrates more profound influence of nanoparticles onto the modulus $l$ and smaller influence on moduli $m$ and $n$. This deviation can be explained by the approximation of homogeneous distribution, which is used in the theory and becomes rough at the nanometer scale for polymer materials.
The observed slight discrepancy of theoretical and experimental values can also be due to unknown real parameters of filler particles. The values taken from \cite{Pabst-2013,Jean-2011} can be considered in our case as estimates only.

\section{Conclusions}

The observed response of nonlinear elastic parameters to addition of different filler particles to polystyrene matrix demonstrates substantial influence of nanofillers onto nonlinear properties of polymer-matrix composites. As shown the linear elastic moduli are affected by such structural changes of the material to considerably lesser extent than nonlinear ones.  The variation of some nonlinear moduli can reach almost 100\% due to the presence of nanofillers in studied composites.

It was also demonstrated that the introduced parameter $\gamma$, besides of being measured in a relatively simple way, seems to be more sensitive to changes of material nonlinearity than the known parameter $\beta$, especially for polymers and polymer composites in which all the  nonlinear elastic moduli usually have the same sign.

For nanocomposites with spherical inclusions (PS+SiO${}_2$ and PS+CB), the nonlinear modulus $l$ and the parameter $\gamma$ demonstrated the most prominent change, especially for PS+CB (more than 80\%). The absolute values of the nonlinear modulus $m$ and the parameter $\beta$ also increased in nanocomposites. At the same time, the variation of the modulus $n$ was less significant.

The nonlinear moduli obtained for nanocomposites with spherical inclusions are in a qualitative agreement with the theoretical predictions calculated using the nonlinear elasticity theory. However, the experiment revealed that the largest nonlinear modulus $l$ demonstrates higher sensitivity to the presence of nanoinclusions than that predicted by the theory. 

For nanocomposites with one-dimensional inclusions (PS+HNT), the variations of nonlinear moduli was qualitatively different. The deviation of the modulus $l$ was opposite and not so significant as for composites with spherical nanoparticles. The parameter $\gamma$ also had an opposite change and became close to zero. At the same time, the modulus $n$ increased by almost 100\%. 

%In spite of comparable concentrations of the nanofillers used, variations of both linear and nonlinear elastic properties observed in PS+SiO${}_2$ composite are less than those in PS+HNT and, especially, in PS+CB composites. That can be explained by relatively low mass of dispersed SiO${}_2$ particles (ca 6 g/L). To provide noticeable effect they have to be added at high volume concentrations, which cause high aggregation and formation of defects in the material.  Interestingly the HNT and CB particles cause opposite changes in $l,n$ moduli and parameter $\gamma$.

%It was also demonstrated that the introduced parameter $\gamma$, besides of being measured in a relatively simple way, seems to be more sensitive to  changes of material nonlinearity than the known parameter $\beta$, especially for polymers and polymer composites in which all the  nonlinear elastic moduli usually have the same sign.

%The nonlinear moduli of nanocomposites with spherical inclusions (PS+SiO${}_2$ and PS+CB) are in quantitative agreement with the theoretical predictions calculated using the nonlinear elasticity theory. However, the experiment demonstrates that the largest nonlinear modulus $l$ demonstrates higher sensitivity to the presence of nanoinclusions than that predicted by the theory. 

\begin{table*}
\caption{\label{tab:table3}Second- and third-order moduli of PS-based nanocomposites obtained from ultrasonic measurements and theoretical modeling.}
%\begin{ruledtabular}
\begin{tabular}{cccccccc}
\textbf{Material}   & $\lambda$,      & $\mu$,          & $l$,            & $m$,            & $n$,            & $\beta$         & $\gamma$\\
                    & GPa             & GPa             & GPa             & GPa             & GPa             &                 &         \\
\hline
PS \cite{Hughes-Kelly1953}
                   & $2.89\pm0.01$ & $1.38\pm0.01$ & $-18.9\pm3.2$ & $-13.3\pm2.9$ & $-10  \pm1.4$ & $ -6.6\pm1.2$ & $ 1.4\pm1.2$ \\
PS commercial      & $2.80\pm0.02$ & $1.44\pm0.01$ & $-46.2\pm1.8$ & $-14.8\pm0.7$ & $ -7.5\pm0.6$ & $-11.8\pm0.4$ & $-2.9\pm0.4$\\
PS pure            & $2.76\pm0.02$ & $1.45\pm0.01$ & $-44.5\pm1.3$ & $-12.8\pm0.5$ & $ -5.7\pm0.4$ & $-10.9\pm0.3$ & $-3.4\pm0.3$\\
%PS pure Univer     & $2.85\pm0.02$ & $1.43\pm0.01$ & $-38.15\pm1.66$ & $-14.66\pm0.65$ & $-6.48\pm0.47$ & $-10.32\\
%
% STANDARD DENSITY:
%PS+20\%SiO${}_2$   & $3.00\pm0.02$ & $1.42\pm0.01$ & $-57.5\pm1.7$ & $-14.2\pm0.8$ & $ -6.5\pm0.6$ & $-13.2\pm0.4$ & $-5.0\pm0.3$\\
%PS+10\%HNT         & $2.84\pm0.02$ & $1.53\pm0.01$ & $-38.5\pm2.0$ & $-17.5\pm0.9$ & $-10.3\pm0.8$ & $-11.0\pm0.5$ & $-0.6\pm0.5$\\
%PS+20\%CB          & $3.26\pm0.02$ & $1.56\pm0.01$ & $-74.7\pm2.5$ & $-17.2\pm0.7$ & $ -4.9\pm0.7$ & $-15.6\pm0.5$ & $-6.3\pm0.5$ \\
%
% CALCULATED DENSITY:
PS+20\%SiO${}_2$ & $3.35\pm0.02$ & $1.58\pm0.01$ & $-64.2\pm1.9$ & $-15.8\pm0.9$ & $ -7.3\pm0.7$ & $-13.2\pm0.4$ & $-5.0\pm0.3$\\
PS+10\%HNT       & $3.02\pm0.02$ & $1.62\pm0.01$ & $-40.9\pm2.1$ & $-18.6\pm1.0$ & $-10.9\pm0.9$ & $-11.0\pm0.5$ & $-0.6\pm0.5$\\
PS+20\%CB       & $3.58\pm0.02$ & $1.71\pm0.01$ & $-81.9\pm2.7$ & $-18.9\pm0.8$ & $-5.4 \pm0.8$ & $-15.6\pm0.5$ & $-6.3\pm0.5$\\
PS+20\%SiO${}_2$ theory \cite{Semenov_Beltukov2020}\!
                    & 3.11            & 1.78            & $-54.8$           &  $-17.1$      &  $-8.7$   &  $-11.8$        & $-3.1$ \\
PS+20\%CB theory \cite{Semenov_Beltukov2020}\!
                    & 3.22            & 1.84            & $-57.8$           &  $-18.2$      &  $-9.3 $        &  $-12.2$      & $-3.1$
\end{tabular}
%\end{ruledtabular}
\end{table*}

\section*{Acknowledgments}
The financial support from Russian Science Foundation
under the grant \# 17-72-20201 is gratefully
acknowledged.

Authors are grateful to V.E. Yudin for valuable discussions and for providing equipment for fabrication of composite samples.

\section*{Data Availability}

The raw/processed data required to reproduce these findings cannot be shared at this time as the data also forms part of an ongoing study.

%\appendix
%\section{My Appendix}
%Appendix sections are coded under \verb+\appendix+.

%\verb+\printcredits+ command is used after appendix sections to list 
%author credit taxonomy contribution roles tagged using \verb+\credit+ 
%in frontmatter.

%\printcredits

%% Loading bibliography style file
\bibliographystyle{model1-num-names}
%\bibliographystyle{cas-model2-names}

% Loading bibliography database
\bibliography{refs}

\end{document}